\documentclass[
  journal=pasa,
  manuscript=research-paper, %% or "review"
  year=202X,
  volume=YY,
]{cup-journal}

\PassOptionsToPackage{authoryear}{natbib}
\usepackage{natbib}

\usepackage{amsmath}
\usepackage{amssymb,microtype,siunitx,booktabs}
\usepackage{longtable}
\usepackage{booktabs}
\usepackage{multirow}
\usepackage{graphicx}
\usepackage{caption}
\usepackage{subcaption}

\newcommand{\ms}{\ensuremath{\mathrm{m\,s^{-1}}}}

\newcommand{\kms}{\mbox{km\,s$^{-1}$}}

\usepackage{xcolor}
\usepackage{soul}        
\sethlcolor{yellow} 

\title[HD38973]{HD\,38973~b - a cold Saturn orbiting a Sun-like star}

\author{Adriana Errico}
\affiliation{University of Southern Queensland, Centre for Astrophysics, West Street, Toowoomba, QLD 4350, Australia}
\email[Adriana Errico]{aberrico@gmail.com}
\author{Robert A. Wittenmyer}
\affiliation{University of Southern Queensland, Centre for Astrophysics, West Street, Toowoomba, QLD 4350, Australia}

\author{Jonathan Horner}
\affiliation{University of Southern Queensland, Centre for Astrophysics, West Street, Toowoomba, QLD 4350, Australia}

\author{Brad Carter}
\affiliation{University of Southern Queensland, Centre for Astrophysics, West Street, Toowoomba, QLD 4350, Australia}

\author{Alexander Wallace}
\affiliation{University of Southern Queensland, Centre for Astrophysics, West Street, Toowoomba, QLD 4350, Australia}

\received {dd Mmm YYYY}
\revised  {dd Mmm YYYY}
\accepted {dd Mmm YYYY}
\published{22 September 202X}

\keywords{Exoplanet astronomy (486) ---  Radial velocity (1332) --- Astrometry (80) } 

\begin{document}

\begin{abstract}

We report the detection of a long-period companion to the nearby solar-type star HD\,38973 using precision radial-velocity measurements. The radial-velocity data reveal a coherent Keplerian signal with a period of $\sim$3000~days and moderate eccentricity, yielding a minimum mass in the sub-Jovian regime.

We complement the radial-velocity analysis with astrometric constraints from the \emph{Hipparcos--Gaia} Catalog of Accelerations (HGCA). Although no significant proper-motion anomaly is detected for HD\,38973, the absence of an astrometric signal provides an informative upper limit on the companion mass. By combining the radial-velocity posterior with the HGCA likelihood, we rule out high-mass solutions at low inclinations and derive a robust upper bound on the true companion mass.  We find the best-fitting true mass to be $0.240_{-0.040}^{+0.102}\,M_{\rm Jup}$, on an orbit with period $2733^{+210}_{-190}$ days, making HD\,38973b a likely cold Saturn.  

This study highlights the diagnostic power of astrometric non-detections when combined with precision radial velocities, demonstrating that meaningful constraints on companion masses can be obtained even in the absence of a detected astrometric signal.

\end{abstract}

\section{Introduction}
\label{sec:int}

The discovery of the first planets orbiting other stars, in the 1990s, marked the start of a great astronomical revolution -- the dawn of the Exoplanet era. Where once we had wondered whether the Solar system was unique, we soon learned that planets were ubiquitous, with almost all stars hosting a planetary retinue. The first planets discovered, however, were very different to those known in the Solar system -- giant planets moving on short period orbits around their host stars \citep[which became known as Hot Jupiters, e.g. ][]{mayor1995jupiter,hotjup1,hotjup2}. 

As time has passed and our ability to find and characterise exoplanets has improved, we have discovered an astonishing diversity of planets -- from those that are super-dense \citep[e.g.][]{Dense2,Dense3,Dense1} to others that have extremely low densities \citep[e.g.][]{fluffy1,fluffy3,fluffy2}. We have found planets that move on orbits so extremely elongated that they resemble those of the Solar system's cometary objects \citep[e.g.][]{highe1,highe2,highe3}, and remarkable chains of planets trapped in mutual mean-motion resonance \citep[e.g.][]{MMR3,MMR1,MMR2}. What we have yet to find, however, is a system that can be truly said to be like our own. 

Throughout that time, planetary systems that resemble our own have remained of great interest. The Solar system is the one planetary system we can study in great and intimate detail\footnote{For a detailed overview of the Solar system, we direct the interested reader to \cite{SSRev}, and references therein.}, and so it has become the template for researchers who intend to search for evidence of life around other stars. Our definitions of habitability, which will determine the best targets for the future search for life, are all based on the Solar system \citep[e.g.][]{Kast93,Lam09,HabRev,Kopp13,Kopp14}, and the perceived unique features of the system have been used to argue that life might be scarce in the cosmos \citep[e.g.][]{Laskar93,RareEarth,SternTect}.

The presence of giant planets on long period orbits -- so called `Jupiter analogues' -- has long been considered a key component for a system to be considered kin to the Solar system. Many studies have considered the effect of such giant planets on Earth-like worlds -- from their role influencing the impact rates on telluric planets \citep[e.g.][]{Wetherill94,Wetherill95,FoF1,FoF2,FoF4,FoF3,Graz16}, to their impact on the delivery of volatiles to planets that form interior to the ice-line \citep[e.g.][]{Chyba87,OB95,FN7,FN9,JontiDHpaper,OB14,OB18}, the manner in which they sculpt the rest of their planetary system \citep[e.g.][]{LHB1,LHB2,Wal11}, and even the degree to which they influence the long-term climate stability of terrestrial planets \citep[e.g.][]{Milank20,MilankStephen,PamMilk}.

Unfortunately, Jupiter-analogues remain challenging to find, with only a relatively small number of such planets being discovered to date \citep[e.g.][]{cj1,cj2,cj3,errico2022HD83443,AdrianaPaper2}. Such planets, moving on long period orbits, are very unlikely to be detected by transit surveys. Radial velocity observations of stars can detect Jupiter-mass planets on long-period orbits, but such work requires a decades-long commitment of observational resources, and only a few surveys have been able to carry out such observations of a small number of bright stars \citep[e.g.][]{Survey1,Survey2,Survey3}. Taken together, those surveys suggest that Solar system analogues - with the innermost massive planets located exterior to the iceline - are neither rare nor common, with an occurrence rate for Jupiter analogues around Sun-like stars calculated to be of order 10\% \citep[e.g.][]{fernandes2019, lagrange23, ToastyWit}.

In the coming years, new data from the \textit{Gaia} space observatory holds the potential to solve this problem. Astrometric observations of stars are an ideal means to detect and characterise massive planets on long-period orbits \citep[e.g.][]{astrometry1, astrometry2, astrometry3, astrometry4}. Furthermore, when such observations are combined with data from radial velocity surveys, they can solve one of the underlying problems for both methods -- namely that both astrometry and radial velocity measurements only see one component of a star's true three-dimensional motion through space. By combining astrometric and radial velocity measurements, it becomes possible to determine the true mass of a given planet, along with its orbital inclination, in addition to fully constraining the planet's orbit around its host star. 

In this work, we apply those principles to announce the discovery of a Saturn-mass planet moving on a long period orbit around the star HD 38973. In Section~\ref{sec:RV}, we present the radial velocity data obtained for HD 38973 since 1998, and model that data to demonstrate the presence of a long-period companion with mass comparable to Saturn. Then, in Section~\ref{sec:Astrometry}, we detail our analysis of the existing astrometric data, before describing our joint inference strategy in Section~\ref{sec:JointInfer}. Finally, we discuss our results and draw our conclusions in Section~\ref{sec:Conclusion}.

\section{Radial-velocity data and model}
\label{sec:RV}

HD\,38973 is a Solar-type G0 dwarf star whose salient properties are summarised in Table~\ref{tab:stellar_params}.  Of particular note for our purposes here, it is a slow rotator amenable to precise radial-velocity (RV) measurements (v sin $i<2$\kms; Soto \& Jenkins 2018).  As such, this star has been targeted by multiple Southern Hemisphere precise RV planet-search programmes.   

\begin{table}
\centering
\begin{threeparttable}
\caption{Stellar parameters of HD\,38973. }
\label{tab:stellar_params}
\begin{tabular}{l c c}
\hline
Parameter & Value & Source\\
\hline
Spectral type                 & G0\,V & (1) \\
Mass, $M_\star$ [$M_\odot$]  & 1.071$^{+0.050}_{-0.044}$ & (3) \\
Radius, $R_\star$ [$R_\odot$]& 1.134$\pm$0.014  & (3)\\
Effective temperature, $T_{\rm eff}$ [K] & 6047$\pm$50 & (2)\\
log $g$ & 4.378$\pm$0.051 & (2) \\
Metallicity, [Fe/H] & 0.01$\pm$0.05 & (2) \\
v sin $i$ (\kms) & 1.97$\pm$0.25 & (3) \\
Parallax [mas] & 34.8288$\pm$0.0171 & (4)\\
Distance (pc)  &  28.712$\pm$0.0141 & (4) \\
\hline
\end{tabular}
\begin{tablenotes}
\footnotesize
\item \textit{Notes.} (1) \cite{gray2006}; (2) \cite{perdelwitz24}; (3) \cite{sotojenkins2018} (4) \cite{gaiaEDR3}.
\end{tablenotes}
\end{threeparttable}
\end{table}

HD\,38973 was observed on 48 epochs from 1998 Jan 16 to 2015 Jan 30 as part of the 18-year Anglo-Australian Planet Search \citep[e.g.][]{tinney2001first, aaps2, aaps3}. Doppler radial velocity measurements were obtained with the UCLES echelle spectrograph \citep{diego:90} at the 3.9-metre Anglo-Australian Telescope (AAT). A 1-arcsecond slit delivers a resolving power of $R\sim$45,000. The spectrograph point-spread function was calibrated using an iodine absorption cell temperature-controlled at 60.0$ \pm $0.1$^{\rm{o}}$C. The iodine cell superimposes a forest of narrow absorption lines from 5000 to 6200\,\AA, allowing simultaneous calibration of instrumental drifts as well as a precise wavelength reference \citep{valenti95,butler96}. The resulting radial velocity shift is measured relative to the epoch of the iodine-free ``template'' spectrum. AAT velocities for HD\,38973 (\ref{app:RV_table}, Table~\ref{tab:AATvels}) span 17 years and have a mean internal uncertainty of 2.0\ms.

We also include 45 RVs from the HARPS spectrograph on the ESO 3.6m telescope at La Silla. The data used here were obtained from the HARPS RVBank\footnote{\url{https://exo-restart.com/wp-content/uploads/Stellar_parameters/HARPS_RVBank.html}} which provides RVs corrected for systematic errors, as well as stellar activity indicators derived from the spectra \citep{trifonov2020harps, perdelwitz24}. The HARPS data span 9 years, from 2003 Oct 29 to 2012 Dec 31, and have a mean internal uncertainty of 0.9\ms. 

\begin{figure}
    \centering
        \includegraphics[width=\linewidth]{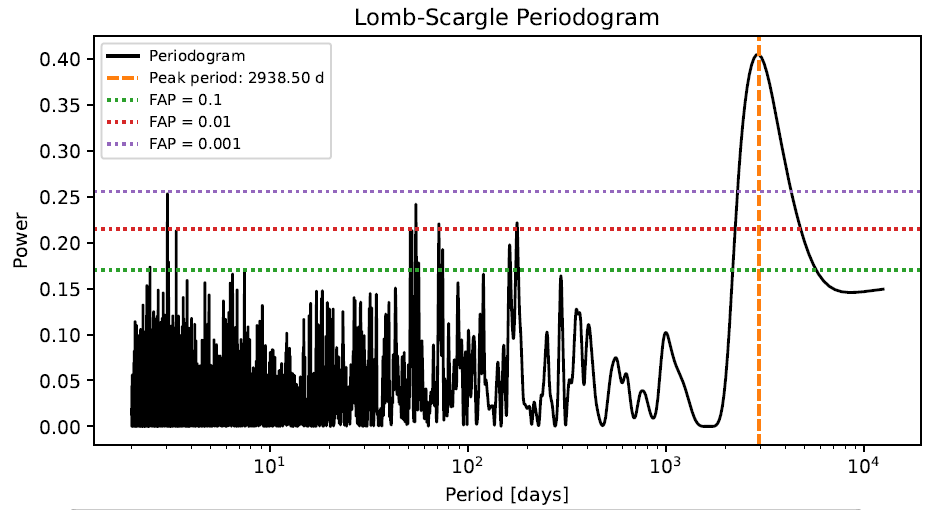}
    \caption{GLS periodogram analysis of the AAT and HARPS radial velocities for HD\,38973.  A highly significant peak is evident at $P\sim$2938 days, motivating the detailed orbital and astrometric investigation presented here. }
    \label{fig:periodo}
\end{figure}

A Generalised Lomb-Scargle (GLS) periodogram of these two data sets revealed a highly significant peak near 3000 days with a false-alarm probability $FAP=6.077\times10^{-8}$ (Figure~\ref{fig:periodo}).  Motivated by this signal, we first carried out a preliminary orbit fit analysis using our own code, implementing a Markov Chain Monte Carlo (MCMC) approach to sample the posterior distribution of the orbital parameters. The RVs are modeled using a single-planet Keplerian of the form:
\begin{equation}
v(t) = K \left[\cos\left(\nu(t) + \omega\right) + e \cos \omega \right] + \gamma_{\rm inst},
\end{equation}
where $P$ is the orbital period, $T_0$ the time of periastron passage, $e$ the eccentricity, $\omega$ the argument of periastron, and $K$ the RV semi-amplitude. Independent velocity offsets and jitter terms are included for each instrument. The AAT data feature a small ($\sim$ 10\ms) upward velocity offset that occurs near JD 2455500, which has been noted in previous analysis of other stars from that survey \citep{Li2024}. The origin of the shift remains unsolved, but as it affects more than half of the 200 Anglo-Australian Planet Search targets, with the same direction and similar magnitude, the cause is almost certainly not astrophysical. We therefore treat the AAT data as two separate sets (``pre'' and ``post'') with an independent offset for each.  

\begin{figure}
    \centering
        \includegraphics[width=\linewidth]{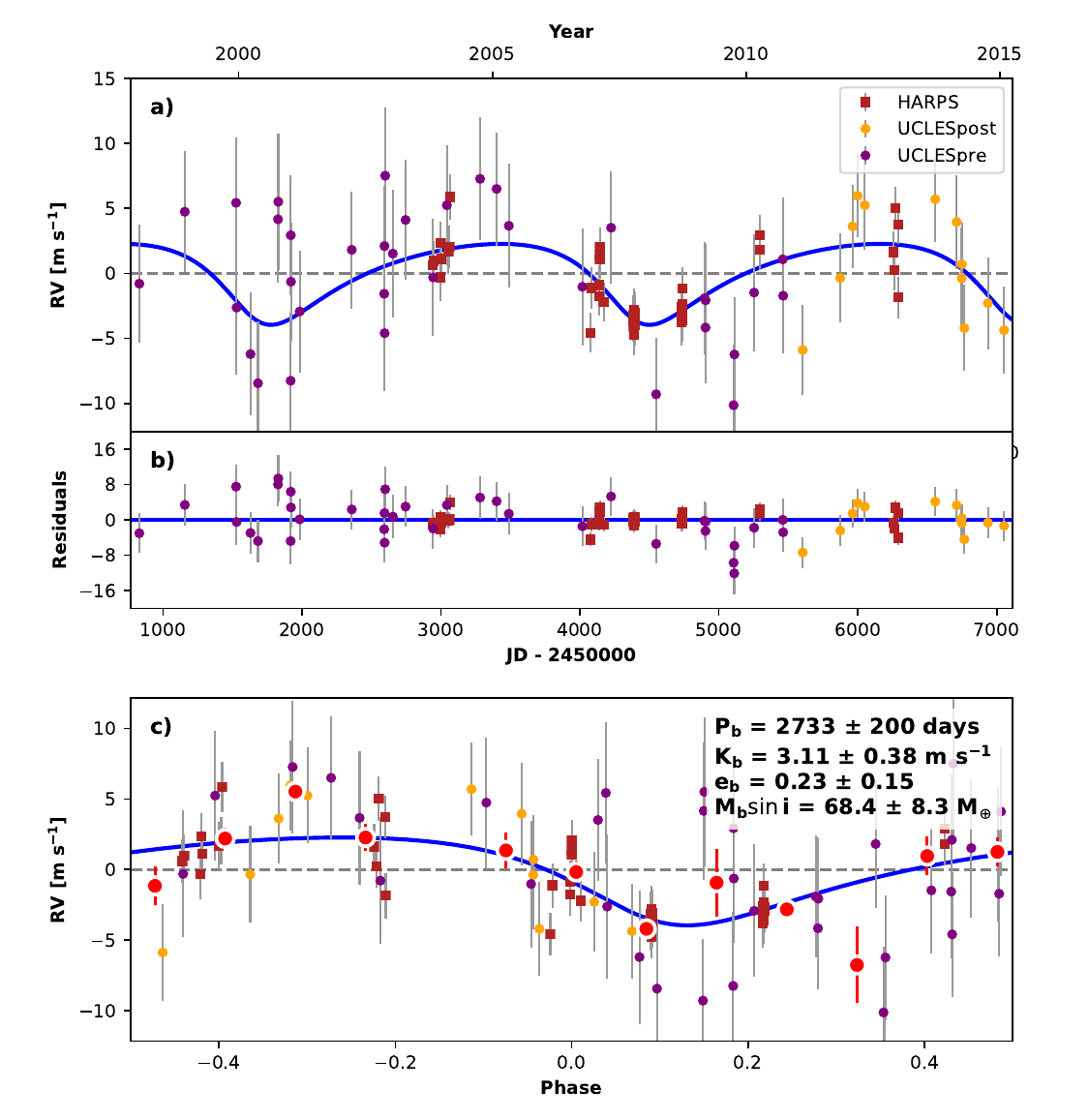}
    \caption{Radial-velocity analysis of HD\,38973. 
a) Best-fit 1-planet Keplerian orbital model for HD\,38973. The maximum likelihood model is plotted while the orbital parameters listed in Table 2 are the median values of the posterior distributions. The thin blue line is the best fit 1-planet model. We add in quadrature the RV jitter term(s) listed in Table 2 with the measurement uncertainties for all RVs. b) Residuals to the best fit 1-planet model. c) RVs phase-folded to the ephemeris of planet b. The small point colours and symbols are the same as in panel a. Red circles are the same velocities binned in 0.08 units of orbital phase.  The phase-folded model for planet b is shown as the blue line.}
    \label{fig:radvel}
\end{figure}

The results of that initial orbital fit were then used as starting values for a more detailed analysis using \texttt{RadVel} \citep{fulton2018radvel}.  The uncertainties derived from the initial fit MCMC posteriors were used to apply gentle Gaussian priors on P, $T_0$, and K, with widths equal to 5 times the $1\sigma$ uncertainties derived previously. We show the results in Table \ref{tab:radvel_params} and complete corner plots are shown in the \ref{app:RVfigs}. The adopted \texttt{RadVel} model fit and residuals are shown in Figure~\ref{fig:radvel}.  

% Radvel results table finalised 13 Feb with table 1's stellar mass.
\begin{table}
\centering
\caption{Orbital parameters of the companion inferred from the combined HARPS and AAT radial velocity data using the adopted \texttt{RadVel} fit.  Quoted uncertainties correspond to the 16th and 84th percentiles of the posterior distributions.}
\label{tab:radvel_params}
\begin{tabular}{l c}
\hline
Parameter & Value \\
\hline
Orbital period, $P$ [days] 
& $2733^{+210}_{-190}$ \\

Time of periastron passage, $T_0$ [JD] 
& $2457085^{+370}_{-400}$ \\

Eccentricity, $e$ 
& $0.23\pm0.15$ \\

Argument of periastron, $\omega$ [deg] 
& $152^{+41}_{-46}$ \\

RV semi-amplitude, $K$ [$\mathrm{m\,s^{-1}}$] 
& $3.11^{+0.39}_{-0.37}$ \\

Systemic velocity (AAT-pre), $\gamma_{\rm AAT}$ [$\mathrm{m\,s^{-1}}$] 
& $-3.79^{+0.82}_{-0.84}$ \\

Systemic velocity (AAT-post), $\gamma_{\rm AAT}$ [$\mathrm{m\,s^{-1}}$] 
& $7.2^{+1.4}_{-1.3}$ \\

Systemic velocity (HARPS), $\gamma_{\rm HARPS}$ [$\mathrm{m\,s^{-1}}$] 
& $0.87^{+0.31}_{-0.32}$ \\

Minimum mass, $M_p \sin i$ [$M_{\rm Jup}$] 
& $0.22\pm0.03$ \\

Semimajor axis, $a$ [au] 
& 3.91$^{+0.21}_{-0.19}$ \\

\hline
\end{tabular}
\end{table}

Considering the stellar mass of $M_\star = 1.071^{+0.050}_{-0.044} M_\odot$ \citep{sotojenkins2018}, the inferred minimum mass of the companion is $M_p \sin i \simeq 0.22\,M_{\rm Jup}$, placing it well within the planetary mass regime. 

The value of $M \textrm{sin} i$ of 0.22$\pm$0.03 is agnostic of the orbital inclination of the planet. Based on the RVs, it is impossible to rule out any particular inclination for the orbit - meaning that it could effectively be anywhere from edge-on to face-on. To obtain an estimate and uncertainty for the true mass of the planet, we draw  inclinations for the system from a uniform distribution between zero and 90$^\circ$, and for each drawn inclination, rescale this minimum mass.  The result of this exercise is detailed further in Section~\ref{sec:JointInfer} where we incorporate astrometric information to place a constraint on the orbital inclination and hence true mass of HD\,38973b. The result of this analysis for the RV data alone is a distribution of true masses, with a peak at 0.242$^{+0.140}_{-0.043}$ $M_{\textrm Jup}$.

\subsection{Analysis of Stellar Activity Indicators}

\begin{figure*}[!t]
\centering

\begin{subfigure}{0.45\textwidth}
\centering
\includegraphics[width=\linewidth]{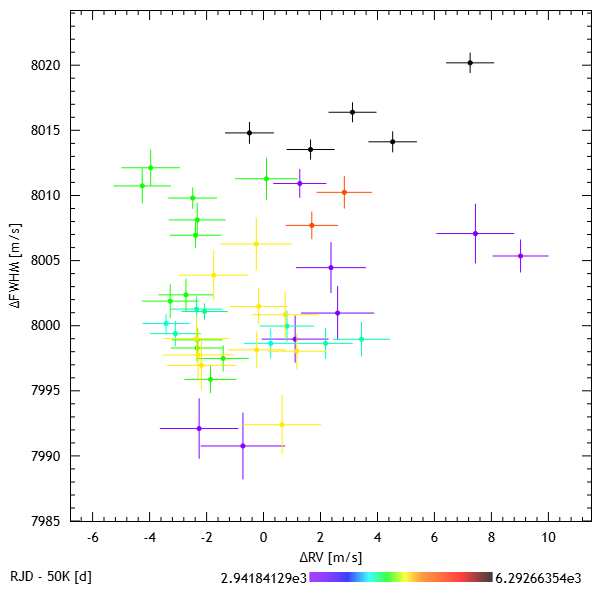}
\caption{CCF-FWHM}
\label{fig:corr-FWHM}
\end{subfigure}
\hfill
\begin{subfigure}{0.45\textwidth}
\centering
\includegraphics[width=\linewidth]{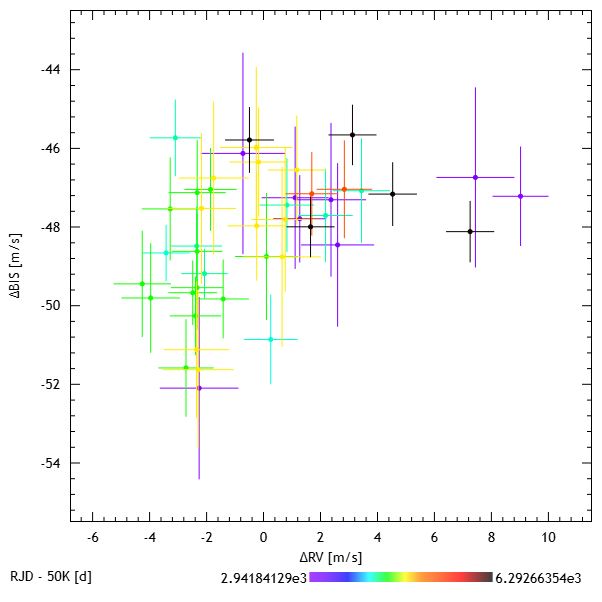}
\caption{CCF-Bisector}
\label{fig:corr-bis}
\end{subfigure}

\vspace{0.5em}

\begin{subfigure}{0.45\textwidth}
\centering
\includegraphics[width=\linewidth]{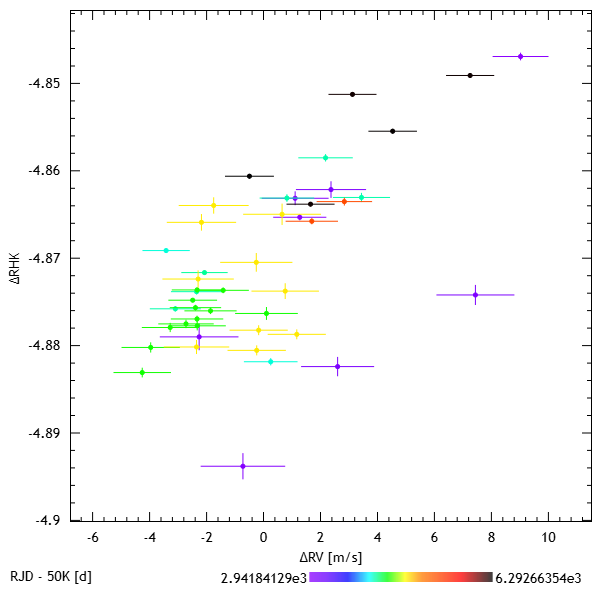}
\caption{$\log R'_{\mathrm{HK}}$}
\label{fig:corr-log}
\end{subfigure}
\hfill
\begin{subfigure}{0.45\textwidth}
\centering
\includegraphics[width=\linewidth]{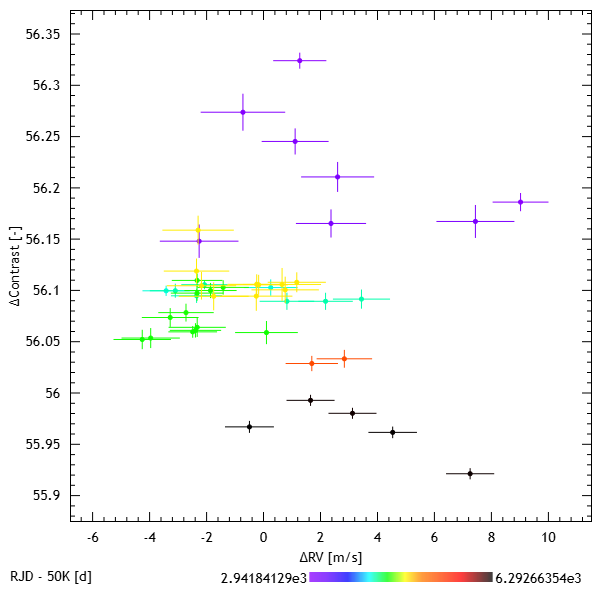}
\caption{CCF-Contrast}
\label{fig:corr-berv}
\end{subfigure}

\caption{Correlations between the radial-velocity measurements and four main stellar activity indicators: (a) FWHM, (b) CCF bisector, (c) $\log R'_{\mathrm{HK}}$, and (d) CCF-Contrast. The colours of the points relate to the dates of observation. No significant correlations are observed in any of the panels.}
\label{fig:correlations}
\end{figure*}

The period and amplitude of the radial-velocity signal we have identified here is worryingly close to the typical influences of long-term stellar magnetic cycles; a notable example being the $\sim$11-year Solar cycle \citep[][e.g.]{lindegren2003, meunier2010}.  We therefore investigate several common activity indicators, which are included for the HARPS spectra.  Figure \ref{fig:correlations} shows the correlations between the radial-velocity measurements and the main stellar activity indicators (FWHM, BIS, $\log R'_{\mathrm{HK}}$, and Contrast). Long-term magnetic cycles are known to induce radial-velocity signals that may mimic the presence of a planetary companion \citep{santos2010do}.  An analysis of HARPS Ca II H\&K chromospheric activity measurements revealed no evidence of a magnetic activity cycle in HD\,38973 \citep{lovis2011harps}.  Consistently, we find no significant correlations between the radial velocities and any of the activity indicators, suggesting that the observed signal is unlikely to be driven by stellar magnetic variability.

% Correlations from DACE without ESPRESSO points

\section{Astrometric Analysis with Hipparcos-Gaia}
\label{sec:Astrometry}
While the radial-velocity analysis provides strong evidence for a long-period companion orbiting HD\,38973, and tightly constrains $P$, $e$, $\omega$, and $K$, this technique alone is insensitive to the orbital inclination and therefore yields only a minimum mass estimate.
Astrometric observations with \emph{Hipparcos} and \emph{Gaia} can help us resolve the mass-inclination degeneracy and assess whether the companion's true mass is compatible with the planetary regime.
\subsection{Astrometric observables}
As a preliminary step, we estimated the expected astrometric signal induced by the companion using the orbital parameters derived from the radial-velocity analysis. For a calculated planetary semi-major axis of $a_p = 3.866\,\mathrm{AU}$, and mass equal to the minimum mass given in Table~\ref{tab:radvel_params}, the corresponding stellar reflex motion has an astrometric semi-major axis of $\alpha \simeq 0.027\,\mathrm{mas}$. 
This value is below the typical single-epoch astrometric precision of both \emph{Hipparcos} and \emph{Gaia}, and therefore a direct estimate of orbital parameters is not expected.  This is also confirmed by the source's low re-normalised unit weight error (RUWE) value of $\sim$1.09, below the accepted threshold of companion detectability \citep{penoyre22,castro24}.
For this reason, we instead estimate an upper limit of the companion's mass, in an attempt to confirm its planetary nature.  Due to the companion's long period of $\sim$8\,years, we focus on the proper motion anomaly (difference between the short-term \emph{Gaia} proper motion and the long-term proper motion inferred from the \emph{Hipparcos--Gaia} baseline) that are sensitive to long-term accelerations \citep{Brandt_2018, Brandt_2021}.  The observed proper motion anomaly is calculated by the long-term baseline proper motion (given by \texttt{pmra\_hg} and \texttt{pmdec\_hg}) and \textit{Gaia}'s proper motion (given by \texttt{pmra\_gaia} and \texttt{pmdec\_gaia}) with cross-calibration and nonlinear corrections applied.  For HD\,38973, this is equal to
\begin{equation}
\Delta \mu_{\alpha^\ast,\mathrm{obs}} = -0.0026\pm0.0344\,\mathrm{mas/yr},
\end{equation}
\begin{equation}
\Delta \mu_{\delta,\mathrm{obs}} = 0.0135\pm0.0383\,\mathrm{mas/yr}.
\end{equation}
\subsection{Astrometric model}
For a given set of orbital parameters, the expected $\boldsymbol{\mu}$ is computed using a forward model of the stellar reflex motion induced by the companion. We first simulate photocentre offsets in the scanning direction, and fit proper motion with a method analogous to \textit{Gaia}'s Astrometric Global Iterative Solution (AGIS) \citep{lindegren2012}.  The observation times and scanning angles are obtained from the \textit{Gaia} Observation Forecast Tool (GOST).  The joint proper motion between the two catalogues is given by
\begin{equation}
    \Delta\boldsymbol{\mu}_{HG} = \frac{\boldsymbol{x}(t_{G})-\boldsymbol{x}(t_{H})}{t_{G}-t_{H}}
    \label{eq:pmhg}
\end{equation}
where ${x}(t_{G})$ and $\boldsymbol{x}(t_{H})$ are the offsets of the photocentre from the barycentre at the \textit{Gaia} and \textit{Hipparcos} reference epochs (2016.0 and 1991.25 respectively).  The proper motion anomaly is then modelled by
\begin{equation}
\Delta\boldsymbol{\mu}_{\rm{model}} = \Delta\boldsymbol{\mu}_{G}-\Delta\boldsymbol{\mu}_{HG}
\end{equation}
From this model, the expected proper-motion anomaly is compared directly with $\Delta\boldsymbol{\mu}_{\mathrm{obs}}$ through the HGCA covariance matrix.

The astrometric likelihood is given by
\begin{equation}
\ln \mathcal{L}_{\rm ast} =
-\frac{1}{2}
\left(
\Delta\boldsymbol{\mu}_{\rm obs} -
\Delta\boldsymbol{\mu}_{\rm model}
\right)^{\!\top}
\mathbf{C}_{\Delta\mu}^{-1}
\left(
\Delta\boldsymbol{\mu}_{\rm obs} -
\Delta\boldsymbol{\mu}_{\rm model}
\right),
\end{equation}
where $\mathbf{C}_{\Delta\mu}$ is the HGCA covariance matrix.
\subsection{Astrometric constraints from HGCA}

Using the HGCA proper-motion anomaly, we evaluated the astrometric constraints on the companion mass implied by the radial-velocity solution.  For a fixed orbital period, eccentricity, argument and time of periapsis inferred from RVs, we computed the expected proper-motion anomaly as a function of companion mass and orbital inclination. For each mass and inclination, the likelihood was marginalised over the longitude of the ascending node by calculating
\begin{equation}
    \ln\mathcal{L}_{\rm total} = \ln\int\limits\mathcal{L}(\Omega)d\Omega.
\end{equation}
\begin{figure}
    \centering
    \includegraphics[width=\linewidth]{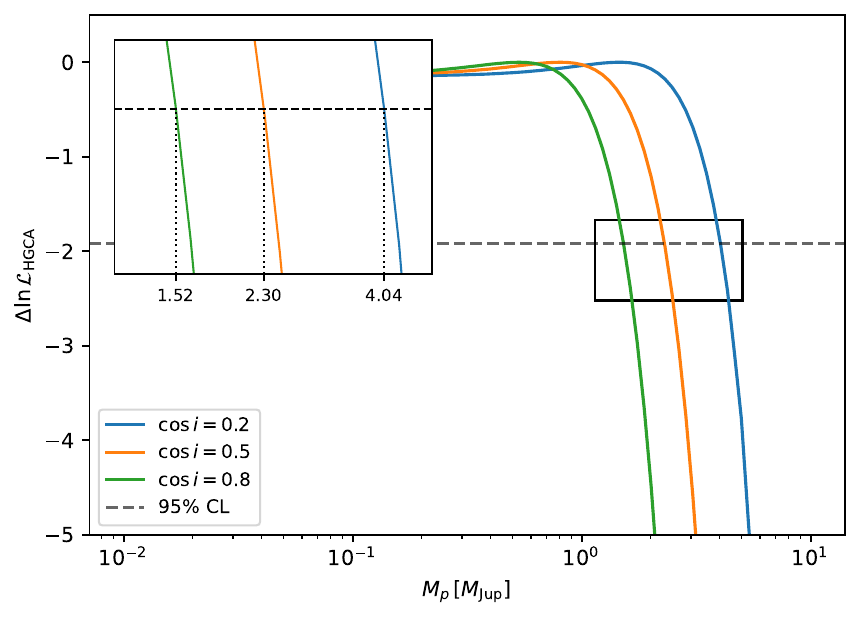}
    \caption{Normalised HGCA log-likelihood as a function of companion mass for HD\,38973, computed for several fixed values of the orbital inclination. For each mass and inclination, the likelihood is marginalised over the longitude of the ascending node. The horizontal dashed line indicates $\Delta\ln\mathcal{L}_{\rm HGCA}=-1.92$, approximately corresponding to a 95\% confidence upper limit on the companion mass.}
    \label{fig:HGCA_mass_constrain}
    \hfill
\end{figure}

Figure \ref{fig:HGCA_mass_constrain} shows the HGCA log-likelihood as a function of companion mass for three\footnote{cos $i$ = [0.2, 0.5, 0.8]} representative values of the orbital inclination. 
For each inclination, the likelihood curve decreases monotonically with increasing mass, reflecting the fact that more massive companions would induce larger proper-motion anomalies that are increasingly inconsistent with the HGCA measurements. 
The horizontal dashed line marks the threshold $\Delta \ln \mathcal{L} = -1.92$, corresponding to a one-sided 95\% confidence level for a single parameter.
The intersection between each likelihood curve and this threshold defines the corresponding 95\% upper limit on the companion mass for that specific inclination. 
These intersections therefore provide inclination-dependent upper limits on $M_p$, illustrating how the astrometric constraints strengthen as the orbit approaches a face-on configuration.\\
These results are expected to improve with the upcoming release of DR4, in late 2026, which will add a new baseline to the HGCA, even though the stellar reflex motion will still be too small for direct detection of this companion.  For this study, we simulate a `minimum astrometric offset' planet an edge-on orbit with mass equal to the minimum value obtained by RV.  Eccentricity and argument/time of periastron were also taken from RV results.  The longitude of the ascending node set to 0$^{\circ}$.  From this, the expected proper motion anomaly was calculated by the same method as for DR3: obtaining the observation epochs from GOST, calculating the effect of the companion over DR4's time window, and subtracting the Hipparcos-Gaia combined proper motion (Equation~\ref{eq:pmhg}).  For DR4, we use an epoch time ($t_{G}$) of 2017.5.  This process was repeated for DR3, and the likelihood functions were combined to obtain a tighter constraint on mass.  We assume the same astrometric uncertainties HGCA covariance matrix for DR3 and DR4.  In reality, DR4 will most likely contain smaller uncertainties, so this can be considered a `worst case scenario.'\\
This simulation was repeated for the DR3 baseline and the likelihoods were combined to obtain the best possible mass limit.  The resultant upper mass limits from just using DR3 and Hipparcos (Figure~\ref{fig:HGCA_mass_constrain}) and adding in DR4 are shown in Table~\ref{tab:astro_mass}.
\begin{table}
\centering
\caption{Astrometric upper mass limits as a function of inclination using proper motion anomaly with DR3 and DR3+DR4.}
\label{tab:astro_mass}
\begin{tabular}{l cc}
\hline
 & \multicolumn{2}{c}{Mass ($M_{\rm Jup}$)} \\
$\cos i$ & DR3+HIP & DR3+DR4+HIP \\
\hline
0.2 & 4.04 & 3.52 \\
0.5 & 2.33 & 3.45 \\
0.8 & 1.52 & 2.98 \\
\hline
\end{tabular}
\end{table}
These results strongly suggest that, the astrometric signal is consistent with a planet and not a brown dwarf or low mass stellar companion.
\subsection{DR2-DR3 Proper Motion Anomaly}
We investigated the possibility of also applying \textit{Gaia}'s `internal' proper motion anomaly between DR2 and DR3, as this would be an ideal time baseline for this companion's period.  However, after applying frame rotation calibration from \citep{Feng24} we found the magnitude of this anomaly to be $\sim$0.2\,mas/yr.  A 0.22\,M$_{\mathrm{J}}$ planet on an edge-on orbit would produce a proper motion anomaly magnitude of $\sim$0.006\,mas/yr.  In order to produce the observed magnitude, the system would need to be almost face-on ($i<4^{\circ}$) which would cause an average RUWE of $\sim$1.3, which is above the companion detection threshold for DR3 of 1.25 \citep{penoyre22,castro24}, and thus significantly higher than the catalogue value of 1.09.\\
A survey of 160 sources of comparable magnitude ($m_{G}\in[5.5,7]$), colour ($Bp-Rp\in[0.6,0.9]$), and RUWE (<1.25) found that the proper motion anomaly is typical for this stellar sample.  We therefore, conclude that the DR2-DR3 proper motion anomaly for this source is more consistent with systematic noise than a substellar companion.\\

\section{Joint inference strategy}
\label{sec:JointInfer}
In order to place a more definitive constraint on the planet's mass, we perform a joint analysis, combining our RV and astrometric analyses.\\
Starting from the posterior samples of the radial-velocity solution, which constrain the minimum companion mass $M_p \sin i$, we generated a distribution of true companion masses by marginalizing over the unknown orbital inclination. Assuming an isotropic distribution of orbital orientations, we drew $\cos i$ values from a uniform distribution in the interval $(0,1)$ and computed the corresponding $\sin i$ for each realization. The true mass for each sample was then obtained as 
\begin{equation}
\label{eq:mass}
M_p = \frac{M_p\sin i}{\sin i}.
\end{equation}
This procedure produces a broad prior distribution for the true companion mass, reflecting the geometric degeneracy inherent to radial-velocity measurements. These values illustrate the strong impact of the inclination prior on the inferred companion mass and motivate the use of external constraints, such as astrometric information, to further restrict the allowed mass range.

These summary statistics provide a compact description of the range of astrometric constraints allowed by unknown orbital orientation and are used in the following section when combining the HGCA information with the radial-velocity results.

To combine the radial-velocity and astrometric information, we adopt a Monte Carlo approach based on the posterior samples obtained from the RV-only MCMC analysis.
For each RV posterior sample, corresponding to a value of $M_p\sin i$, an orbital inclination is randomly drawn assuming an isotropic orientation prior.  For this, we assume a uniform distribution of $\cos i\in [0,1]$.  We ignore negative values of $\cos i$ (clockwise orbits) because, due to the small astrometric signal, this reversal of on-sky motion will have a negligible effect on likelihood.
This allows us to convert the minimum mass into a true companion mass via eq. \ref{eq:mass}

For each realization, the expected proper-motion anomaly is computed and compared with the HGCA measurements by evaluating the astrometric log-likelihood marginalised over the longitude of the ascending node.
These likelihood values are then used to assign statistical weights to each Monte Carlo sample, effectively downweighting orbital configurations that are incompatible with the observed proper-motion anomaly.

A weighted resampling of the Monte Carlo ensemble is performed using these astrometric likelihoods, yielding a posterior distribution for the companion mass that incorporates both the RV constraints and the HGCA non-detection.

\begin{figure*}
    \centering
    
    \begin{subfigure}{0.46\textwidth}
        \centering
        \includegraphics[width=\linewidth]{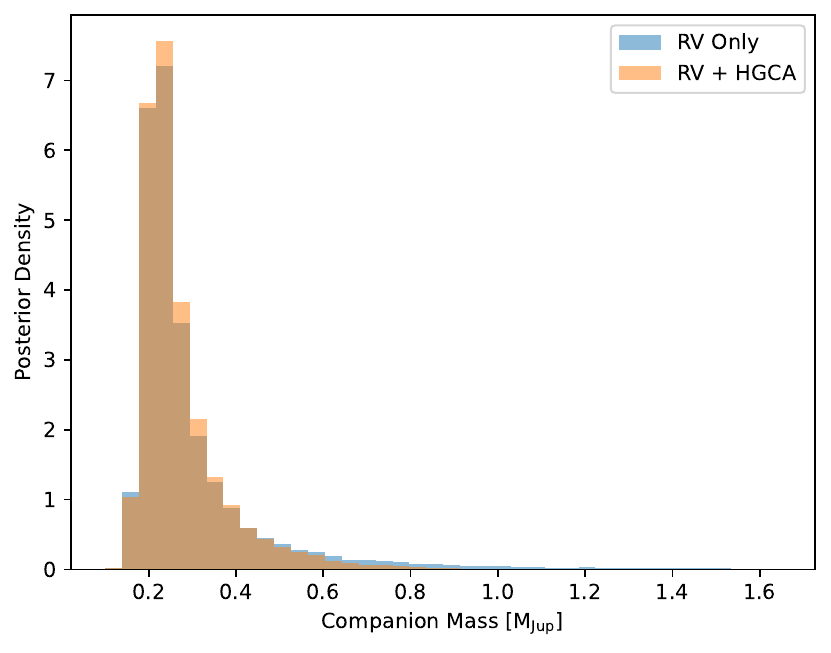}
        \caption{Linear scale.}
        \label{fig:mass_linear}
    \end{subfigure}
    \hfill
    \begin{subfigure}{0.48\textwidth}
        \centering
        \includegraphics[width=\linewidth]{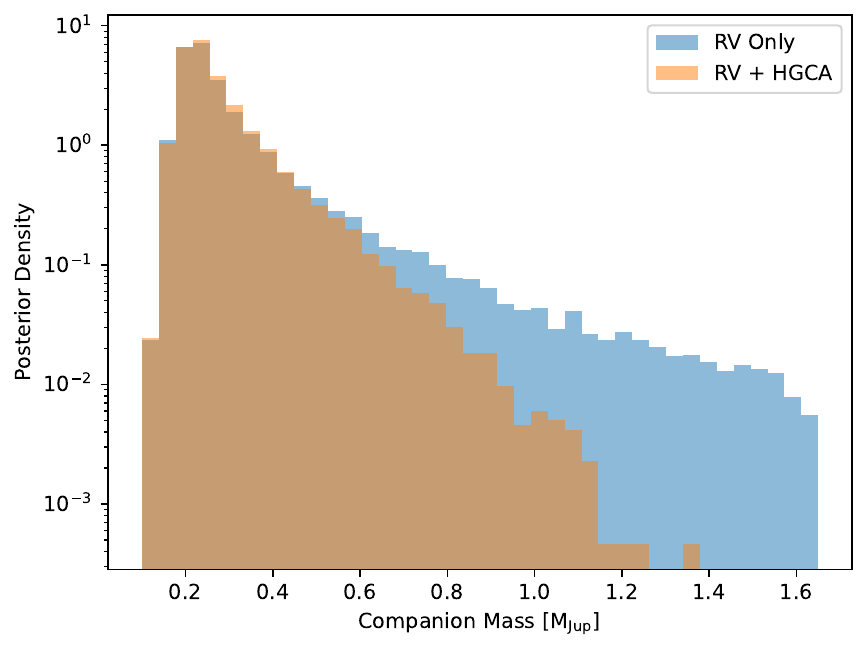}
        \caption{Logarithmic scale on the vertical axis.}
        \label{fig:mass_log}
    \end{subfigure}
    
    \caption{Posterior mass distributions obtained from the RV-only and joint RV and astrometric analyses. The left-hand panel shows the likelihood of different planet masses on a linear scale, with the results from the RV measurements alone illustrated in blue, and those from our joint analysis shown in orange. The right panel presents the same data but with the likelihood given on a logarithmic scale. This latter panel shows the significant impact that including the astrometric non-detection has on the tail of possible true masses that result from the sin~$i$ ambiguity inherent to RV observations.}
    \label{fig:mass_comparison}
\end{figure*}

Figure~\ref{fig:mass_comparison} presents the posterior mass distributions obtained from the radial velocity (RV) analysis alone, and from the joint RV+HGCA solution. The left panel shows the distributions on a linear scale, facilitating a direct comparison of the median values and highest-density regions. The right panel displays the same information using a logarithmic scale on the vertical axis, highlighting differences in the tails of the distributions, and emphasising the region in which the combined analysis leads to significantly greater constraints on the planetary mass. While the RV-only solution exhibits a pronounced high-mass tail driven by the inclination degeneracy, the inclusion of astrometric constraints significantly suppresses this behaviour, yielding a more tightly constrained companion mass distribution.

From the Monte Carlo realization of the RV posterior assuming isotropic orbital orientations, we find the mass distribution peaks at $0.242_{-0.043}^{+0.138}\,M_{\rm Jup}$, with a $95\%$ upper limit of $0.662\,M_{\rm Jup}$.  However, if we incorporate the astrometric information from the HGCA via likelihood weighting, we attain a peak mass of $0.240_{-0.040}^{+0.102}\,M_{\rm Jup}$ with a $95\%$ upper limit of $0.484\,M_{\rm Jup}$.  This demonstrates how the small astrometric signal can be used to further constrain the true mass and inclination.  Specifically, these results place a minimum value on sin $i$ of 0.5 at a $95\%$ confidence level, which constrains the inclination to be between 27$^{\circ}$ and 153$^{\circ}$, and strongly rules out near-face-on orbital solutions.
In the following section we place these results in a broader context, discuss their implications for the nature of the proposed companion to HD\,38973, and summarise the main conclusions of this work.

\section{Discussion and conclusions}
\label{sec:Conclusion}

We report the detection of a long-period radial-velocity signal in HD\,38973 consistent with a sub-Jovian planetary companion on a multi-year orbit ($P \sim 2733$ days). Independent Keplerian fits using HARPS and AAT data yield mutually consistent orbital solutions, supporting the robustness of the RV detection. The inferred minimum mass ($M_{\textrm RV}\sim0.22\pm0.03~M_{\textrm Jup}$) places the companion well within the planetary regime.

We explored the astrometric detectability of this signal using the \emph{Hipparcos--Gaia} proper-motion anomaly formalism. The expected astrometric wobble induced by the RV solution is of order $\sim$0.027~mas, which lies below the nominal sensitivity of both \emph{Hipparcos} and \emph{Gaia} for a single target. Consistent with this expectation, the measured HGCA proper-motion anomaly for HD\,38973 is statistically consistent with zero.

% RW: I rearranged the text here to address the referee confusion about the really real adopted mass. 

Despite the absence of a significant astrometric detection, the HGCA data provide meaningful constraints on the companion mass. By marginalizing over the unknown longitude of the ascending node and exploring a range of orbital inclinations, we derive inclination-dependent astrometric likelihoods that exclude high-mass companions incompatible with the observed proper-motion anomaly. We further combined the astrometric information with the RV-derived posterior samples through importance reweighting. This analysis yields a 95\% upper limit of 0.484 M$_{\textrm Jup}$ on the true companion mass that is significantly tighter than the RV-only constraint of 0.662 M$_{\textrm Jup}$), despite the limited astrometric sensitivity in this specific case. It is clear that the combined RV+HGCA analysis robustly rules out stellar or brown-dwarf companions and confirms the planetary nature of the signal.

These mass upper limits show a strong dependence on inclination, as expected for long-period systems. The resultant best fit planetary mass based on our joint inference strategy for HD~38973~b is $0.240_{-0.040}^{+0.102}\,M_{\rm Jup}$.  We adopt this solution as our final, true mass determination, establishing HD\,38973b as a cold sub-Saturn mass planet moving on a $\sim$7.5-year orbit.

Even when the expected signal lies below the detection threshold, proper-motion anomalies can exclude large regions of parameter space and place physically meaningful constraints on companion mass and inclination. As \emph{Gaia} astrometry continues to improve, similar analyses will become increasingly valuable for the characterisation of long-period planets detected by radial velocities.

Our constraint on the orbital inclination of HD~38973~b also slightly increases the calculated transit probability. Assuming an isotropic distribution of inclinations, the transit probability for a planet orbiting a Sun-like star at a distance of 3.91 au is $\sim$0.15\,\%. For HD~38973~b, our analysis rules out orbits that are perfectly face on, and therefore increases this to 0.20\,\%.  Whilst this is only a minor change in transit probability in this case, this does demonstrate the capacity of such analysis, in the future, to help identify systems with a realistic potential for transits to be observed, through the combined use of RV and astrometric observations.

\clearpage

\appendix

\section{RV data from the AAT}
\label{app:RV_table}

\begin{table}[H]
\centering
\caption{AAT/UCLES Radial Velocities.  Data after BJD 2455500 are treated as coming from a separate instrument with an independent velocity offset. }
\label{tab:AATvels}
\begin{tabular}{lrr}
\hline
BJD & Velocity (\ms) & Uncertainty(\ms) \\
\hline
2450829.99918  &      -4.5  &    1.7  \\
2451157.15853  &       1.0  &    2.1  \\
2451526.08642  &       1.7  &    2.8  \\
2451530.13642  &      -6.4  &    3.0  \\
2451629.88997  &      -9.9  &    2.2  \\
2451683.84560  &     -12.2  &    2.3  \\
2451828.19373  &       0.4  &    2.5  \\
2451830.03484  &       1.8  &    3.2  \\
2451919.09343  &     -12.0  &    3.2  \\
2451920.05674  &      -0.8  &    1.9  \\
2451921.14349  &      -4.4  &    1.9  \\
2451983.92757  &      -6.7  &    2.2  \\
2452358.89067  &      -1.9  &    1.7  \\
2452592.13514  &      -5.3  &    2.3  \\
2452594.20596  &      -1.6  &    2.0  \\
2452595.16603  &      -8.3  &    1.6  \\
2452599.17832  &       3.8  &    3.2  \\
2452654.10593  &      -2.2  &    2.6  \\
2452745.90806  &       0.4  &    2.0  \\
2452944.25819  &      -4.0  &    1.7  \\
2453043.02349  &       1.5  &    2.0  \\
2453282.28818  &       3.5  &    2.2  \\
2453401.02595  &       2.8  &    1.3  \\
2453488.85716  &      -0.1  &    2.3  \\
2454018.27510  &      -4.8  &    1.7  \\
2454224.86679  &      -0.2  &    1.2  \\
2454548.96825  &     -13.0  &    1.3  \\
2454897.02433  &      -5.7  &    1.2  \\
2454903.96035  &      -7.9  &    1.1  \\
2454904.96273  &      -5.8  &    1.3  \\
2455106.26032  &     -13.9  &    2.1  \\
2455111.16160  &     -16.2  &    2.4  \\
2455112.20570  &     -10.0  &    1.5  \\
2455253.99715  &      -5.2  &    1.7  \\
2455460.28709  &      -2.7  &    2.3  \\
2455463.29140  &      -5.5  &    1.6  \\
\hline 
2455604.06644  &       1.3  &    1.9  \\
2455874.22384  &       6.8  &    1.9  \\
2455962.03340  &      10.8  &    1.4  \\
2455998.93372  &      13.1  &    1.4  \\
2456050.87700  &      12.4  &    1.9  \\
2456557.29393  &      12.9  &    1.6  \\
2456711.99005  &      11.1  &    2.2  \\
2456747.91185  &       6.8  &    2.6  \\
2456748.91059  &       7.9  &    1.4  \\
2456766.88788  &       3.0  &    1.7  \\
2456936.26497  &       4.9  &    2.1  \\
2457053.05800  &       2.8  &    1.8  \\
\hline
\end{tabular}
\end{table}

\section{Additional figures}
\label{app:RVfigs}

% Primera figura: ocupa 2 columnas
\begin{figure*}[!t]
    \centering
    \includegraphics[width=1\textwidth]{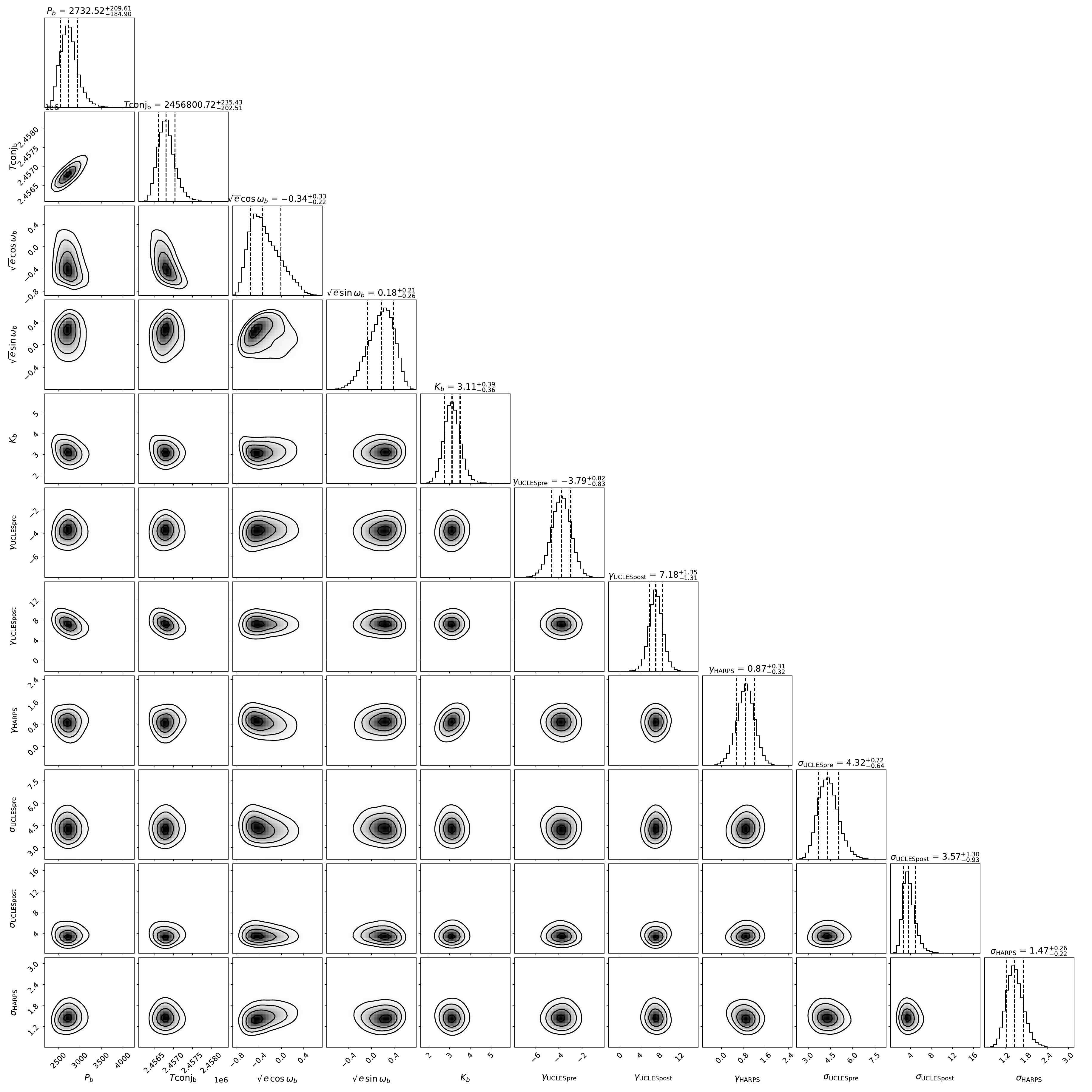}
    \caption{Corner plot of posteriors from the \texttt{RadVel} model fit. All parameters exhibit unimodal posteriors.}
    \label{fig:radvelcornerplot}
\end{figure*}

% Segunda figura: una columna
\begin{figure}[h]
    \centering
    \includegraphics[width=\linewidth]{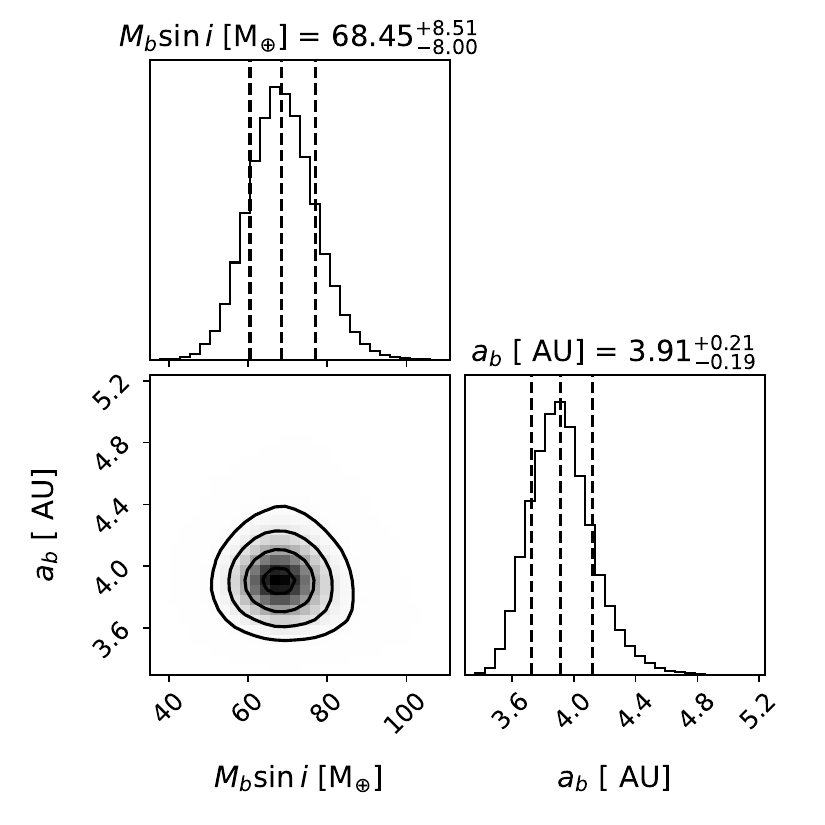}
    \caption{Corner plot of MCMC posteriors for the derived planetary parameters $a$ and $M_p \sin i$.}
    \label{fig:derivedparams}
\end{figure}

\clearpage

\bibliography{bib}

\end{document}